\documentclass[aps,prl,twocolumn,groupedaddress,amsmath,longbibliography,superscriptaddress]{revtex4-2}

\usepackage[final]{changes}

\usepackage{graphicx}% Include figure files
\usepackage{float}

\usepackage{xcolor,graphicx}
\usepackage{xspace}
\usepackage{amsmath}
\usepackage{braket}
\usepackage{hyperref}
\usepackage{upgreek}

\newcommand{\up}{\ket{\uparrow}\xspace}
\newcommand{\down}{\ket{\downarrow}\xspace}
\newcommand{\aux}{\ket{\mathrm{aux}}\xspace}

\newcommand{\Be}{\textsuperscript{9}Be\textsuperscript{+}\xspace}
\newcommand{\qcpl}{\ket{\uparrow}_1\bra{\downarrow}_1+\ket{\uparrow}_2\bra{\downarrow}_2}
\newcommand{\singlet}{\left(\ket{\uparrow\downarrow}-\ket{\downarrow\uparrow}\right)/\sqrt{2}}
\newcommand{\triplet}{\left(\ket{\uparrow\downarrow}+\ket{\downarrow\uparrow}\right)/\sqrt{2})}
\newcommand{\Slevel}{\textsuperscript{2}S\textsubscript{1/2}\xspace}
\newcommand{\Plevel}{\textsuperscript{2}P\textsubscript{1/2}\xspace}

\begin{document}

\author{Daniel C. Cole}
\affiliation{National Institute of Standards and Technology, 325 Broadway, Boulder, CO 80305, USA}

\author{Stephen D. Erickson}
\affiliation{National Institute of Standards and Technology, 325 Broadway, Boulder, CO 80305, USA}
\affiliation{Department of Physics, University of Colorado, Boulder, CO 80309, USA}

\author{Giorgio Zarantonello}
\affiliation{National Institute of Standards and Technology, 325 Broadway, Boulder, CO 80305, USA}
\affiliation{Department of Physics, University of Colorado, Boulder, CO 80309, USA}

\author{Karl P. Horn}
\affiliation{Theoretische Physik, Universit\"at Kassel, Heinrich-Plett-Stra{\ss}e 40, 34132 Kassel, Germany}

\author{Pan-Yu Hou}
\affiliation{National Institute of Standards and Technology, 325 Broadway, Boulder, CO 80305, USA}
\affiliation{Department of Physics, University of Colorado, Boulder, CO 80309, USA}

\author{Jenny J. Wu}
\affiliation{National Institute of Standards and Technology, 325 Broadway, Boulder, CO 80305, USA}
\affiliation{Department of Physics, University of Colorado, Boulder, CO 80309, USA}

\author{Daniel H. Slichter}
\affiliation{National Institute of Standards and Technology, 325 Broadway, Boulder, CO 80305, USA}

\author{Florentin Reiter}
\affiliation{Institute for Quantum Electronics, ETH Z\"{u}rich, Otto-Stern-Weg 1, 8093 Z\"{u}rich, Switzerland}

\author{Christiane P. Koch}
\affiliation{Theoretische Physik, Universit\"at Kassel, Heinrich-Plett-Stra{\ss}e 40, 34132 Kassel, Germany}
\affiliation{Dahlem Center for Complex Quantum Systems and Fachbereich Physik, Freie Universit\"at Berlin, Arnimallee 14, 14195 Berlin, Germany}

\author{Dietrich Leibfried}
\email{dietrich.leibfried@nist.gov}
\affiliation{National Institute of Standards and Technology, 325 Broadway, Boulder, CO 80305, USA}

\title{Resource-efficient dissipative entanglement of two trapped-ion qubits}

\begin{abstract}
 We demonstrate a simplified method for dissipative generation of an entangled state of two trapped-ion qubits. Our implementation produces its target state faster and with higher fidelity than previous demonstrations of dissipative entanglement generation and eliminates the need for auxiliary ions. The entangled singlet state is generated in $\sim$7 ms with a fidelity of 0.949(4).  The dominant source of infidelity is photon scattering. We discuss this error source and strategies for its mitigation.
\end{abstract}

\maketitle

Engineered dissipation has potential as a powerful tool for quantum applications~\cite{Poyatos1996,Verstraete2009}. Dissipation may be used for preparation of non-classical states, including entangled states, and this approach can have reduced sensitivity to certain common experimental imperfections and limitations~\cite{Kastoryano2011, Morigi2015}. Unlike unitary approaches, dissipative dynamics can produce desired target states from unknown or uncontrolled input states; examples in atomic physics include laser cooling and optical pumping. Further, some dissipative protocols can be implemented by continuous, stationary control fields, and can therefore be applied to prepare and continuously stabilize entangled states in the presence of noise. Numerous protocols for dissipative preparation of non-classical states have been demonstrated~\cite{Krauter2011, Barreiro2011, Lin2013, Kienzler2015, Shankar2013, Kimchi-Schwartz2016, Liu2016}, and still more have been proposed and explored~\cite{Plenio1999,Morigi2015, Kastoryano2011,Carr2013,Rao2013,Ticozzi2014,Reiter2016,Shao2017,Bentley2014,Horn2018,Doucet2020,Cole2021}. An important characteristic of initial demonstrations~\cite{Lin2013, Shankar2013} was the use of strong driving fields to create resonances that were resolved and addressed by weaker drives~\cite{Vacanti2009, Kastoryano2011, Reiter2012a}. These weaker drives could populate the target state without providing a path out of it in the limit where the timescales for the strong dressing drive and the weaker addressing drives were well-separated. Recently, schemes have been proposed that avoid these timescale hierarchies. Instead, these schemes make more efficient use of experimental resources such as symmetries and auxiliary degrees of freedom~\cite{Bentley2014, Horn2018, Doucet2020, Cole2021, Malinowski2021}, and are generally expected to produce the desired target state with higher fidelity in less time.

Horn {\it et al.}\ have proposed a protocol for dissipative generation of an entangled singlet state $\ket{S}=\singlet$ of two trapped-ion qubits~\cite{Horn2018}. This scheme improves upon the demonstration in Ref.~\cite{Lin2013} by eliminating the timescale hierarchy and the need for sympathetic cooling, thereby reducing the required number of ions from four to two. In addition to qubit levels $\up$ and $\down$, the protocol uses a stable auxiliary level $\ket{\mathrm{aux}}$ and a short-lived excited state $\ket{e}$, along with a mode of collective motion of the ions. In their proposal, Horn {\it et al.}\ applied quantum optimal control to explore the limits of this scheme, predicting singlet fidelities above 0.98 in the case that heating of the motional mode used for the protocol could be kept low. An important fundamental source of heating is recoil of the ions after photon scattering. The recoil heating rate is linked to the strengths of the interactions that generate the singlet state. In this Letter, we employ this protocol to generate an entangled singlet state with $\sim$~0.95 fidelity, limited by photon scattering errors including recoil heating. We discuss how photon scattering limits the singlet fidelity, theoretically investigate the large-Raman-detuning limit, and present strategies for improving the performance of the protocol.

\begin{figure}[t]
    \begin{centering}
	\includegraphics[]{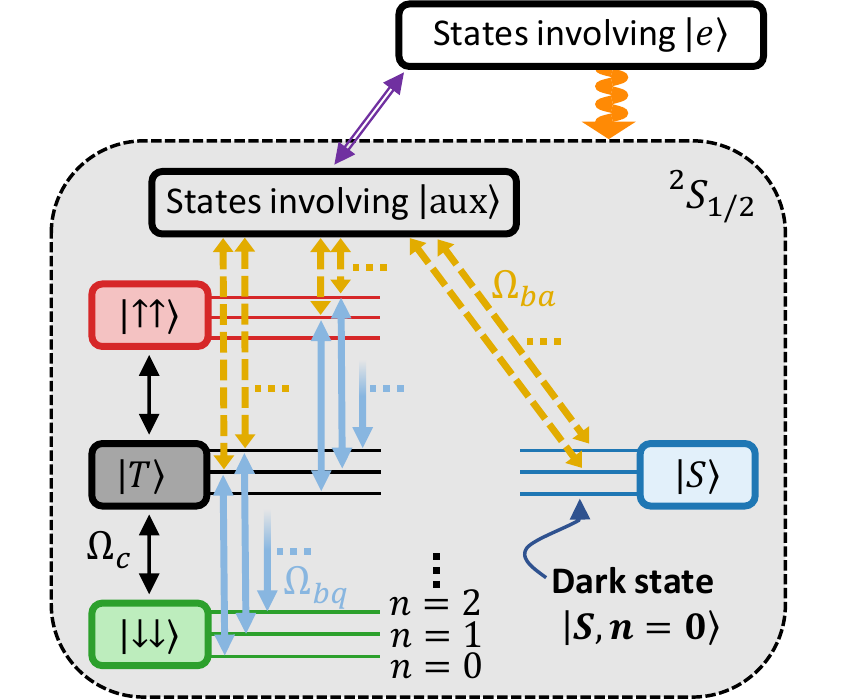}
	\caption{Protocol for dissipative singlet generation. Four interactions combine to generate the target state $\ket{S,n=0}$ in the joint Hilbert space of two ions and their collective motion. Blue-sideband transitions (anti-Jaynes-Cummings interactions) are depicted by solid blue and dashed yellow arrows, and a qubit carrier interaction (implementing the identity on the motion) is depicted by thin black arrows. This carrier interaction is required to depopulate the $\ket{\uparrow\uparrow,n=0}$ state, which is otherwise dark. Excitation of the single-ion $\ket{\mathrm{aux}}$ state to $\ket{e}$ is shown by a double purple arrow, and decay from $\ket{e}$ back to the $^2 S_{1/2}$ ground state is shown by the snaking orange line. The two-qubit basis states are shown with colors corresponding to those in Fig. \ref{fig:singlet_gen}. Next to each qubit state are the first few rungs of the motional number state ladder, and ellipses indicate continuation of interactions to higher number states. No path exists out of the state $\ket{S,n=0}$, which is populated by decay from states involving $\ket{e}$. \label{fig:scheme}}
	\end{centering}
\end{figure}

The concept for the protocol is shown in Fig.~\ref{fig:scheme}. It involves simultaneous application of four global interactions, of which three are unitary: blue-sideband (anti-Jaynes-Cummings) couplings $\ket{\downarrow, n}\leftrightarrow\ket{\uparrow, n+1}$ and $\ket{\mathrm{aux},n}\leftrightarrow\ket{\uparrow,n+1}$ driven by Hamiltonians $H_{bq}$ and $H_{ba}$, respectively, and a qubit carrier transition $\ket{\downarrow}\leftrightarrow\ket{\uparrow}$ driven by Hamiltonian $H_c$. The states $\ket{n}$ are number states of the motional degree of freedom with creation operator $a^\dag$. The Hamiltonians are:
\begin{eqnarray}
H_{bq} = \frac{\hbar\Omega_{bq}}{2}a^\dag\left(\vphantom{\frac{1}{2}}\qcpl\right) + H.c., \\
H_{ba}=\frac{\hbar\Omega_{ba}}{2}a^\dag\left(\vphantom{\frac{1}{2}}\ket{\uparrow}_1\bra{\mathrm{aux}}_1+\ket{\uparrow}_2\bra{\mathrm{aux}}_2\right) + H.c., \\
H_c=\frac{\hbar\Omega_c}{2}\left(\vphantom{\frac{1}{2}}\qcpl\right) + H.c., \label{eq:Hc}
\end{eqnarray}
where $H_c$ implements the identity on the motion, the subscripts $1$ and $2$ label the two ions, and $\Omega_I$ denotes the Rabi frequency of interaction $H_{I}$. A fourth interaction provides dissipation in the form of spontaneous transitions from the auxiliary state as $\aux\rightarrow\up$, $\down$, or $\aux$. This is engineered by coupling $\aux$ to $\ket{e}$, which is chosen so that it may only decay to one of these three levels.

As depicted in Fig. \ref{fig:scheme}, the interactions $H_{bq}$ and $H_c$ couple the states $\ket{\downarrow\downarrow}$, $\ket{T}=\triplet$, and $\ket{\uparrow\uparrow}$ within the {total-spin-$1$} qubit manifold and, together with $H_{ba}$, provide a path for one of the qubits to transition to $\aux$ when starting in any of these states, regardless of the initial motional occupation $n$. The dissipative pumping out of $\aux$ then allows population to be continuously reshuffled until it arrives in the joint state $\ket{S, n=0}$. At this point, the population becomes trapped because $\ket{S}$ is invariant under the qubit interactions $H_{c}$ and $H_{bq}$, and coupling of the $\up$ component of $\ket{S}$ to $\aux$ due to $H_{ba}$ only occurs when $n>0$. Neglecting errors and imperfections, the theoretical steady-state fidelity for generation of $\ket{S, n=0}$ is unity.

We realize this protocol with two trapped \Be ions. The ions are confined along the axis of a linear Paul trap~\cite{Blakestad2010}. A combination of static and RF electric potentials at $\sim83$~MHz applied to the trap electrodes confines the ions such that they have an equilibrium spacing along the axis of the trap of $\sim$3.7 $\upmu$m and exhibit quantized collective motion in three dimensions. The frequencies for the in-phase and out-of-phase (`stretch') axial motional modes are 4 MHz and $f_s=$ 7 MHz, respectively, and the stretch mode is used to engineer the entanglement.

\begin{figure}
    \begin{centering}
	\includegraphics[]{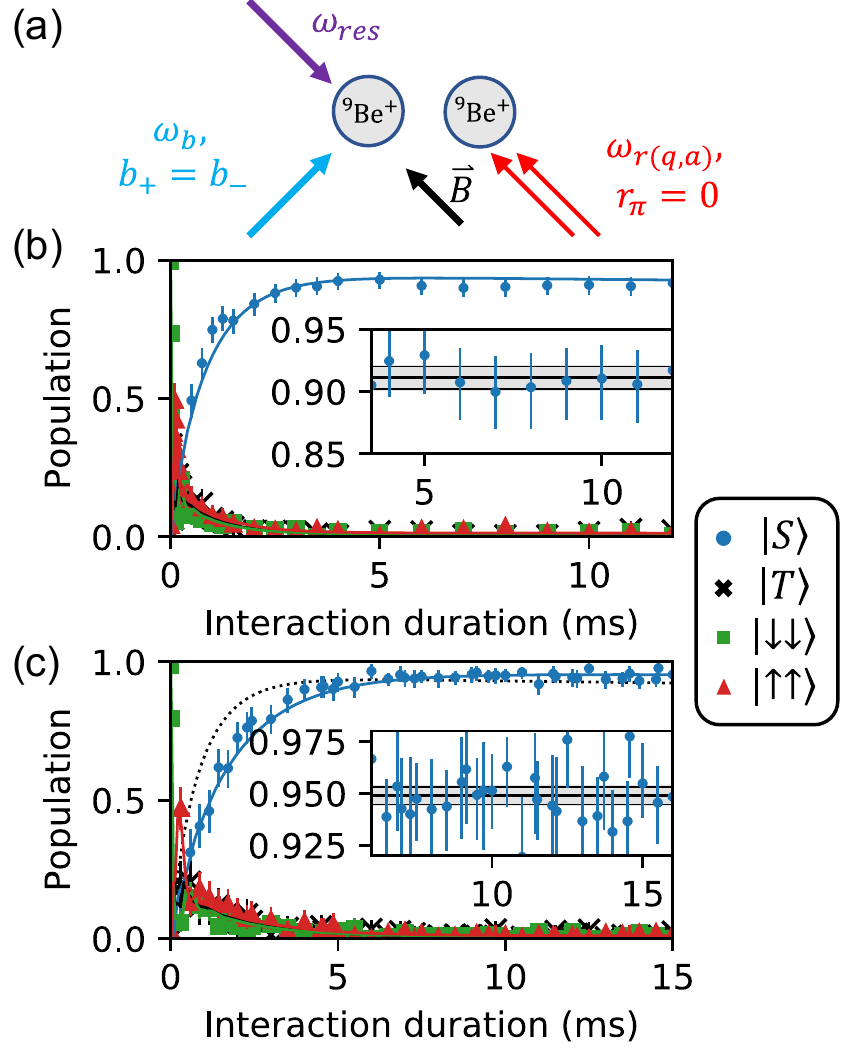}
	\caption{Experimental geometry and results. (a) Trapped ions, magnetic field, and $\vec{k}$ vectors for four laser beams: the higher-frequency Raman beam (blue), two co-propagating lower-frequency Raman beams at frequencies $\omega_{rq}$ and $\omega_{ra}$ (red), and a resonant beam with variable frequency $\omega_{res}$ that drives either the $\ket{\mathrm{aux}}\leftrightarrow\ket{e}$ coupling or the cycling transition. Beams have $\sim$25 $\upmu$m waists and illuminate both ions approximately equally. Constraints on the polarizations of the Raman beams $b$ and $r$, as indicated next to the $\vec{k}$ vectors by components $(b/r)_{\pm,\pi}$ (see text), arise due to their orientations with respect to the quantization field. (b, c) Measured populations in four basis states as a function of interaction duration for Raman detunings of $-$315 GHz (b) and $-$450 GHz (c). Solid lines are simulations with no free parameters. For $-$315 GHz detuning, the simulation includes a $\phi$ error of 0.05 rad and uses the measured $\aux$ depletion time of 34 $\upmu$s. The simulated singlet curve from (b) is replicated in (c) as a dotted black line for comparison. Insets show data on the fidelity plateau. Three horizontal black lines and shading indicating the average fidelity on the plateau and a 95 \% confidence interval generated by bootstrapping, and error bars indicate 95 \% confidence intervals on individual points. \label{fig:singlet_gen}}
	\end{centering}
	
\end{figure}

We apply a $\sim$~11.9 mT magnetic quantization field~\cite{Langer2005} and identify the levels $\down$, $\up$, and $\aux$ with Zeeman sublevels of the \Be $\Slevel$ ground state labelled by hyperfine and magnetic quantum numbers $F$ and $m_F$: $\ket{\downarrow}=\ket{F=2,m_F=2}$, $\ket{\uparrow}=\ket{1,1}$, and $\ket{\mathrm{aux}}=\ket{2,1}$. The Hamiltonian $H_c$ is realized using $\sim$ 1.018 GHz microwave radiation from an external antenna, and the Hamiltonians $H_{bq}$ and $H_{ba}$ are realized by driving stimulated Raman transitions with 313 nm laser radiation tuned hundreds of gigahertz below the $\Slevel\leftrightarrow\Plevel$ transition. The beam geometry is depicted in Fig. \ref{fig:singlet_gen}a. The Raman transitions are driven on the blue motional sideband corresponding to the excitation of the axial stretch mode, which is chosen because it has a lower heating rate than the in-phase axial mode. This is due to its reduced sensitivity to homogeneous electric fields, which arises because the mode eigenvectors for the two ions are exact opposites~\cite{King1998}. Effective decay out of $\aux$ is engineered by driving a unitary coupling between $\aux$ and $\ket{e}=\ket{^2 P_{1/2}, F=2, m_F = 2}$, which decays at a rate $\Gamma\approx 2\pi\times$20 MHz back to $\up$, $\down$, and $\aux$ with approximate branching ratio 5:4:3~\cite{Lin2013}. This coupling is driven resonantly by a 313 nm $\hat{\sigma}_+$-polarized repump laser. Angular momentum conservation dictates that $\ket{e}$ can decay only to one of these three levels, and other transitions that may be driven by the same laser are far off-resonant.

The microwave field, with wavelength $\lambda_{\mu w}\gg|\vec{r}_1-\vec{r}_2|\sim3.7$ $\upmu$m, is nearly the same at the positions $\vec{r}_1$ and $\vec{r}_2$ of the two ions. In the interaction picture for the qubit levels, the Hamiltonian implemented by the microwave radiation can be written in the form given by Eq.~\eqref{eq:Hc}. This defines a relationship between the orientations of the two qubits' Bloch spheres. The qubit sideband interaction then implements the experimental interaction-picture Hamiltonian $H^{(e)}_{bq}$~\cite{Leibfried2003a, Cole2021}:
\begin{align}
H^{(e)}_{bq} = \frac{\hbar\Omega_{bq}}{2}a^\dag&\left(e^{i(\Delta\vec{k}\cdot \vec{r}_1+\theta)}\ket{\uparrow}_1\bra{\downarrow}_1\right.  \nonumber \\ &- \left.e^{i(\Delta\vec{k}\cdot \vec{r}_2+\theta)}\ket{\uparrow}_2\bra{\downarrow}_2\right) + H.c.\\ \nonumber
=e^{i\Phi}\frac{\hbar\Omega_{bq}}{2}&a^\dag\left(\ket{\uparrow}_1\bra{\downarrow}_1 - e^{i\phi}  \ket{\uparrow}_2\bra{\downarrow}_2\right) + H.c. 
\end{align}
Here $\Delta\vec{k}$ is the difference wavevector between the Raman beams, and the sign difference arises because the two ions move in opposite directions in the stretch mode. We have introduced the phases $\phi=\Delta\vec{k}\cdot(\vec{r}_2-\vec{r_1})$ and $\Phi = \Delta\vec{k}\cdot\vec{r}_1+\theta$, where $\theta$ is a reference phase for the interference pattern between the two Raman beams that fluctuates from shot to shot due to lack of interferometric stability between the Raman beams. As a result, the Bloch-sphere rotation axis that is defined by $\Phi$ fluctuates. On the other hand, $\phi$ is stable so long as the vectors $\vec{r}_2-\vec{r}_1$ and $\Delta\vec{k}$ are stable. By setting $\phi$ to $\pi$ as described in the Supplementary Information (SI), $H_{bq}^{(e)}$ is made to coincide with $H_{bq}$ up to the fluctuating rotation axis defined by $\Phi$. These fluctuations have negligible effect on generation or invariance of the singlet because they are slow relative to the entanglement dynamics~\cite{Gaebler2016}.

In order to implement two stimulated-Raman sideband transitions simultaneously, we apply far-detuned laser light at three frequencies $\omega_b$ (higher frequency `blue' beam) and $\omega_{r(q,a)}$ (`red' beams, with subscripts denoting the corresponding Hamiltonian) with frequency differences $\omega_b-\omega_{rq} = (E_\uparrow-E_\downarrow)/\hbar + 2\pi f_s$ and $\omega_b - \omega_{ra} = (E_\uparrow - E_{\mathrm{aux}})/\hbar + 2\pi f_s$, where $E_j$ is the energy of state $j$.  Importantly, in this three-frequency configuration $\omega_{rq}-\omega_{ra} = (E_\downarrow-E_{\mathrm{aux}})/\hbar$ so that the two red beams can resonantly drive the stimulated-Raman $\ket{\downarrow}\leftrightarrow\ket{\mathrm{aux}}$ carrier transition. This would depopulate the singlet state. However, the red beams' $\vec{k}$ vector is approximately parallel to the quantization field. As a result, the component $r_\pi$ of the red beams' polarization unit vector $(r_-, r_\pi, r_+)$, with entries corresponding to $\hat{\sigma}_-$, $\hat{\pi}$, and $\hat{\sigma}_+$ polarizations, is $r_\pi\approx0$. The Rabi frequency of the $\ket{\downarrow}\leftrightarrow\ket{\mathrm{aux}}$ coupling is proportional to this component, so the coupling is strongly suppressed.

We implement this singlet generation protocol and investigate its performance. In principle, the system can be initialized in any mixture of states in which each ion is in $\up$, $\down$, or $\ket{\mathrm{aux}}$ and $n$ is not too large. For increased efficiency and repeatability, we begin by approximately preparing $\ket{\downarrow\downarrow,n=0}$ with optical pumping, Doppler cooling, and sideband cooling. We then simultaneously apply the four interactions for a variable interaction duration $t$. Finally, we measure the populations in four two-qubit basis states by performing global analysis rotation pulses on the two qubits and then performing fluorescence detection on the $\down\leftrightarrow\ket{^2P_{3/2},F=3,m_F=3}$ cycling transition. From the photon count histograms for each condition, maximum-likelihood estimates are obtained for the populations $P_{n,A}(t)$ with $n$ ions in the bright $\down$ state under analysis condition $A$. We use three analysis conditions: no rotation, a $\pi$ pulse, and a $\pi/2$ pulse with randomized phase. These yield the populations $P_{n,I}$, $P_{n,\pi}$, and $P_{n,\pi/2}$, respectively. From these observations, basis-state populations are obtained as~\cite{Lin2013}:
\begin{align}
P_{\downarrow\downarrow} &= P_{2,I}, \label{eq:popdd} \\
P_{\uparrow\uparrow} &= P_{2,\pi}, \\
P_{S} -P_{ll} \equiv X &= 1-2P_{0,\pi/2} - (P_{2,I} + P_{2,\pi})/2, \\
P_{T} &= 2P_{2,\pi/2} - (P_{2,I} + P_{2,\pi})/2. \label{eq:popT}
\end{align}
Formally, the singlet population exceeds $X$ by the population $P_{ll}$ (`leakage-leakage') with both ions in states other than $\{\up,\down\}$. However, this population is very small and $P_S\approx X$ in practice.

We investigate singlet generation for two values of the detuning of the Raman beams from the $\Slevel\leftrightarrow\Plevel$ transition, the importance of which is described below. We show the results in Fig. \ref{fig:singlet_gen}b and c. In each case, we plot measured populations obtained from Eqs.~\eqref{eq:popdd}-\eqref{eq:popT}, along with uncertainties determined from 10,000 bootstrap resamplings of the data. In the inset of each figure we show the data corresponding to a pseudo-steady-state fidelity plateau and a confidence interval (CI) for the plateau fidelity. This CI and the plotted uncertainties are bias-corrected 95 \% bootstrap CIs~\cite{Efron1987}. For a Raman detuning of $-$315 GHz we measure a fidelity (CI) of 0.911 ([0.902, 0.920]), and for $-$450 GHz we measure 0.949 ([0.945, 0.953]). We elaborate on the bootstrapping procedure in the SI.

Figure \ref{fig:singlet_gen}b and c also show simulations of the dynamics as solid and dashed lines. The simulations use the measured Rabi frequencies of the unitary interactions, the depletion time constant of the $\ket{\mathrm{aux}}$ state by the repumper laser, the Lamb-Dicke parameter for the stretch mode, and the Stark shifts induced by the Raman lasers, all determined in separate measurements. The simulations also incorporate spontaneous Raman and Rayleigh scattering driven by the Raman lasers~\cite{Ozeri2007}. Recoil associated with these scattering events and with the repumping transitions is included. Finally, the simulations include a unitary coupling between $\ket{\downarrow}$ and $\ket{\mathrm{aux}}$ arising from a residual non-zero $\hat{\pi}$-polarization component $r_\pi$ of the red Raman beams~(SI). The peak fidelity predicted by the simulation in the $-$450 GHz detuning case is 0.954, consistent with the upper CI bound of 0.953 for the average fidelity between 6 ms and 16 ms. For the $-$315 GHz detuning case the predicted peak fidelity is 0.946. Including an error of 0.05 rad for the phase $\phi$ in Eq.~(4), corresponding to the typical calibration uncertainty, reduces the peak fidelity to 0.935. Simulating the experiment with an $\aux$ repumping time constant of 51 $\upmu$s instead of the measured 34 $\upmu$s brings the predicted peak fidelity to 0.912. The repumper amplitude is not stabilized during the experiment and is known to drift.  We present simulation details in the SI.

\begin{figure}
    \begin{centering}
	\includegraphics[]{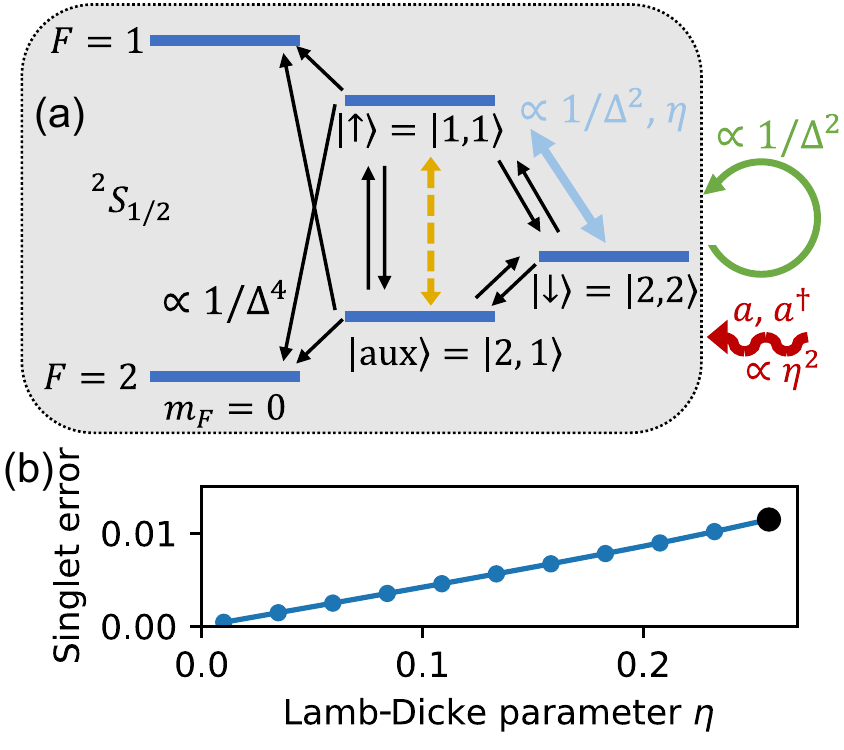}
	\caption{Photon scattering error in singlet generation. (a) A summary of effects associated with scattering of laser photons. These include stimulated Raman sideband transitions (thick light blue arrow and dashed yellow arrow), spontaneous Raman transitions (thin black arrows), and Rayleigh scattering (green loop indicating the identity operation on the internal state of the ions). These processes asymptotically scale with the detuning as $1/\Delta^2$, $1/\Delta^4$, and $1/\Delta^2$, respectively. Recoil leads to heating (modelled by jump operators proportional to products of $a$ and $a^\dag$ and indicated by the snaking red arrow) at a rate proportional to $\eta^2$ to leading order. (b) A calculation of the error in singlet generation as a function of Lamb-Dicke parameter $\eta$ in the large-detuning limit; increasing the strength of the confining potential and therefore decreasing $\eta$ leads to better performance. The larger black dot indicates the value $\eta = 0.257$ used in the experiment. \label{fig:photon_scattering}}
	\end{centering}
\end{figure}

This singlet-generation protocol is robust against a number of typical experimental errors, including magnetic field fluctuations and laser phase noise. On the other hand, the scheme is sensitive to \textit{differential} effects between the two ions, including differences in the Rabi frequencies of the qubit transitions and differential qubit frequency shifts (caused by e.g. magnetic field gradients and differential ac Stark shifts). In our implementation, we have been able to suppress these differential effects so that they are negligible. This is demonstrated by direct measurements of the size of these effects (SI) and also by the agreement of the model with the data. For $-$315 GHz ($-$450 GHz) Raman detuning we calculate an infidelity contribution of 0.008 (0.009) from the residual $\ket{\downarrow}\leftrightarrow\ket{\mathrm{aux}}$ coupling. Calibration errors likely contribute to the infidelity in the $-$315 GHz detuning case as described above. In both cases, the remaining infidelity is due to undesired photon scattering events. 

We depict the relevant stimulated and spontaneous photon scattering processes in Fig. \ref{fig:photon_scattering}a. Spontaneous Raman transitions within the $\up$, $\down$, $\aux$ manifold can be corrected by the singlet-generation dynamics and so do not accumulate, but instead decrease the steady-state fidelity. Spontaneous Raman transitions to leakage states outside this manifold lead to permanent (to first order) population loss, and so lead to fidelity decay. In principle, Rayleigh scattering has two effects: First, Rayleigh scattering can cause decoherence of the qubit. This effect occurs at a rate proportional to the sum of the squared differences between the scattering amplitudes off of the two states for each polarization~\cite{Horn2018, Uys2010}. However, the singlet state is in a decoherence-free subspace~\cite{Lidar1998, Duan1998,Kielpinski2001, Langer2005}, so \textit{differential} decoherence between the two ions is required to affect the singlet fidelity. This occurs only to the extent that the environment resolves which of the two ions scattered a photon~\cite{Eichmann1993}, which is expected to be a small effect for the $\sim$3.7 $\upmu$m-spaced ions. Therefore we neglect Rayleigh decoherence in our model for the experiment. The second effect of Rayleigh scattering is heating of the collective motion due to recoil after scattering events. This heating provides a path out of the target $\ket{S,n=0}$ state, and is included in our model as an important error source.

The infidelity due to spontaneous photon scattering can be reduced at the cost of increased singlet preparation time. Limitations on this approach come from restrictions on the preparation time and timescales at which other errors (e.g. $\ket{S}\leftrightarrow\ket{T}$ coupling due to magnetic field gradients) become relevant. The relative rate of spontaneous Raman transitions can be reduced by increasing the Raman detuning $\Delta$, because the asymptotic scalings of the rates for stimulated and spontaneous Raman scattering are $1/\Delta^2$ and $1/\Delta^4$, respectively. This suggests implementation of the scheme with $|\Delta|$ as large as is practical. In the large-detuning limit $|\Delta|\rightarrow\infty$, the only remaining error source is recoil heating due to Rayleigh scattering (neglecting differential Rayleigh decoherence). We investigate the protocol's performance in this limit by optimizing the laser polarizations and interaction strengths. For the same Lamb-Dicke parameter $\eta=0.257$ used in the experiment, we calculate a fidelity of 0.989 and optimal (respecting the geometric constraints shown in Fig. \ref{fig:singlet_gen}a) Raman beam polarizations of blue-beam $\hat{\pi}$ component $b_\pi = 0.59$ and red-beam $\hat{\sigma}_+$ component $r_+=0.88$. These polarizations are close to the polarizations $b_\pi = 0.62$, $r_+\approx 1$ used in the experiment, chosen to be near-optimal and experimentally convenient.

The stimulated-Raman sideband Rabi rate scales as $\eta$ while the recoil heating rate scales as $\eta^2$, so the error in the large-detuning limit can be reduced by decreasing $\eta$. We numerically investigate the dependence of the steady-state singlet fidelity in the large-detuning limit as a function of $\eta$ and present the results in Fig. \ref{fig:photon_scattering}b. We find that the error decreases linearly with $\eta$ and falls below 0.01 (0.001) at $\eta=0.229$ (0.024). The time to approach the asymptotic fidelity scales as $1/\eta$ due to the reduced Rabi rates for the stimulated Raman sideband transitions.

Another possibility to improve the fidelity may be to incorporate sympathetic cooling. Periods of cooling should be interleaved with periods of the singlet-generation dynamics, since otherwise the cooling interferes with the desired coupling $\ket{\downarrow,n=0}\leftrightarrow\ket{\uparrow,n=1}\leftrightarrow\ket{\mathrm{aux},n=0}$. We find in simulations that if the stretch mode is re-initialized to $n=0$ at intervals equal to the period $2\pi/\Omega_{ba}$ of the $H_{ba}$ coupling, then the fidelity in the large-detuning limit increases to 0.994. However, we also find that without cooling the singlet population has a steady-state motional occupation of $\bar{n}=0.002$. Ground-state cooling performance to at least this level would be required to improve the fidelity. Generally, the steady-state temperature of the singlet state is determined by effects (e.g. recoil) that also limit ground-state cooling, so this kind of strategy may be difficult to productively implement in practice. A final possibility to improve the performance would be driving the sidebands not with Raman lasers but with magnetic field gradients~\cite{Wineland1998, Mintert2001, Ospelkaus2008, Wolk2017, Srinivas2019}. Such interactions typically have smaller sideband Rabi frequencies and would therefore have slower entanglement dynamics, but could make photon scattering error negligible. 

Our demonstration of dissipative singlet generation with fidelity of $\sim$0.95, along with the related work by Malinowski {\it et al.}~\cite{Malinowski2021}, is a step forward in dissipative production of entangled resource states. These works indicate a path towards fidelities that could allow productive incorporation of dissipative protocols into practical trapped-ion platforms for quantum information processing. In this work, the agreement between the photon-scattering error model and the data indicate that numerical simulations can be a powerful tool for optimizing trapped-ion implementations of dissipative protocols in the future, and also supports our conclusion that the current limitation on singlet fidelity arises from photon scattering errors. Our work has further investigated the important role of these errors in entanglement generation, which has been considered in depth for unitary approaches~\cite{Ozeri2007} and represents an outstanding challenge for the realization of practical trapped-ion quantum computers~\cite{Bruzewicz2019a}.

The authors thank Ethan Clements and Shawn Geller for comments on the manuscript and Emanuel Knill and Scott Glancy for helpful discussions. This work was supported by IARPA and the NIST Quantum Information Program.  D. C. C. acknowledges support from a National Research Council postdoctoral fellowship. S. D. E. acknowledges support from the National Science Foundation under grant DGE 1650115. P.-Y. H and J. J. W. acknowledge support from the Professional Research Experience Program (PREP) operated jointly by NIST and University of Colorado Boulder. F. R. acknowledges financial support from the Swiss National Science Foundation (Ambizione grant no. PZ00P2$\_$186040). K. P. H. and C. P. K. acknowledge financial support from the Federal State of Hesse, Germany through the SMolBits project within the LOEWE program.

\bibliography{DissipativeSingletGeneration}

\end{document}

\author{Daniel C. Cole}
\affiliation{National Institute of Standards and Technology, 325 Broadway, Boulder, CO 80305, USA}

\author{Stephen D. Erickson}
\affiliation{National Institute of Standards and Technology, 325 Broadway, Boulder, CO 80305, USA}
\affiliation{Department of Physics, University of Colorado, Boulder, CO 80309, USA}

\author{Giorgio Zarantonello}
\affiliation{National Institute of Standards and Technology, 325 Broadway, Boulder, CO 80305, USA}
\affiliation{Department of Physics, University of Colorado, Boulder, CO 80309, USA}

\author{Karl P. Horn}
\affiliation{Theoretische Physik, Universit\"at Kassel, Heinrich-Plett-Stra{\ss}e 40, 34132 Kassel, Germany}

\author{Pan-Yu Hou}
\affiliation{National Institute of Standards and Technology, 325 Broadway, Boulder, CO 80305, USA}
\affiliation{Department of Physics, University of Colorado, Boulder, CO 80309, USA}

\author{Jenny J. Wu}
\affiliation{National Institute of Standards and Technology, 325 Broadway, Boulder, CO 80305, USA}
\affiliation{Department of Physics, University of Colorado, Boulder, CO 80309, USA}

\author{Daniel H. Slichter}
\affiliation{National Institute of Standards and Technology, 325 Broadway, Boulder, CO 80305, USA}

\author{Florentin Reiter}
\affiliation{Institute for Quantum Electronics, ETH Z\"{u}rich, Otto-Stern-Weg 1, 8093 Z\"{u}rich, Switzerland}

\author{Christiane P. Koch}
\affiliation{Theoretische Physik, Universit\"at Kassel, Heinrich-Plett-Stra{\ss}e 40, 34132 Kassel, Germany}
\affiliation{Dahlem Center for Complex Quantum Systems and Fachbereich Physik, Freie Universit\"at Berlin, Arnimallee 14, 14195 Berlin, Germany}

\author{Dietrich Leibfried}
\email{dietrich.leibfried@nist.gov}
\affiliation{National Institute of Standards and Technology, 325 Broadway, Boulder, CO 80305, USA}

\title{Supplementary Information for ``Resource-efficient dissipative entanglement of two trapped-ion qubits''}

\setcounter{figure}{3}

\maketitle

\section{Numerical simulations of dissipative dynamics}
\subsection{Simulation method}
We use the Python QuTiP package \cite{Johansson2013} to compute evolution of the density matrix $\rho$ according to the Lindblad master equation:
\begin{equation}
\partial_t \rho = -i[H,\rho] +\mathcal{L}_D\rho,
\end{equation}
where $H$ is the Hamiltonian and $\mathcal{L}_D$ is the Lindblad dissipator:
\begin{align}
\mathcal{L}_D \rho &= \sum_j \mathcal{L}_j\rho \\
 &= \sum_j \left[L_j \rho L_j^\dagger - \frac{1}{2}\left\{\rho, L_j^\dagger L_j\right\}\right].
\end{align}
Here $\mathcal{L}_j$ are the full dissipator contributions for each of the jump operators $L_j$ that represent deliberate controlled dissipation and also undesirable dissipation mechanisms.  

To include spontaneous scattering in the model, the intensities and polarizations of the Raman beams are deduced by measuring the ac Stark shifts they induce on the ions and comparing to calculations according to second-order perturbation theory \cite{Wineland1998, Wineland2003, Leibfried2003a}. Recoil after scattering must then be incorporated into the model. In the Schr\"{o}dinger picture for the motion and interaction picture for the qubit levels, recoil associated with transition matrix element $m_j$ contributes jump operators of the form~\cite{Leibfried2003a, Ozeri2007}:
\begin{equation}
    L_j = e^{i \left(k_{z,L} - k_{z,s}\right)z} m_j,
\end{equation}
where $m_j = \sqrt{\gamma}\ket{f}\bra{i}$ is a rate-scaled matrix element for transition $\ket{i}\rightarrow \ket{f}$ for initial and final internal states $\ket{i}$ and $\ket{f}$ with rate $\gamma$, $k_{z,L}$ and $k_{z,s}$ are the axial projections of the wavevectors for the incident laser photon and the scattered photon, respectively, and $z$ is the operator for the position along the trap axis of the ion in question.

For our experimental geometry as described in the main text, for scattering of a Raman photon, and considering the effect of the recoil on the stretch mode with annihilation operator $a$, the full Lindblad dissipator term for matrix element $m_j$ including recoil is \cite{Leibfried2003a}:
\begin{equation}
    \mathcal{L}_j = - \frac{1}{2} \left\{m_j^\dagger m_j, \rho\right\} + \int d\theta d\phi P_\epsilon(\theta,\phi) m_j\mathcal{I}_Sm_j^\dag \label{eq:intdissipator}
\end{equation}
with
\begin{equation}
    \mathcal{I}_S = e^{i\eta\left((1 - \sqrt{2}u(\theta,\phi)\right)(a+a^\dagger)} \rho  e^{-i\eta\left((1 - \sqrt{2}u(\theta,\phi)\right)(a+a^\dagger)}. \label{eq:IS}
\end{equation}
Here $P_\epsilon(\theta,\phi)$ is the emission pattern for polarization $\epsilon$ in a spherical coordinate system oriented along the magnetic quantization field and 
\begin{equation}
    u(\theta,\phi) = \left( \sin{\theta}\cos{\phi}-\cos{\theta} \right)/\sqrt{2}.
\end{equation}

Transforming to the interaction picture for the motion, one can write the quantity $\mathcal{I}_I$ corresponding to Schr\"{o}dinger-picture $\mathcal{I}_S$ (Eq. \eqref{eq:IS}) as:
\begin{align}
    \mathcal{I}_I = \sum_{m=0}^\infty \sum_{n=0}^m \sum_{p=n-m}^n  &\left(\frac{(\eta f)^{2m}(-1)^{m+p}}{ n!(n-p)!(m-n)!(m-n+p)!}\times\right. \nonumber\\ a^n a^{\dag(n-p)} & \rho a^{\dag(m-n+p)} a^{(m-n)} \left.\vphantom{\frac{1}{2}}\right) , \label{eq:I_Sp}
\end{align}
where $f = (1- \sqrt{2}u(\theta,\phi))$. It is straightforward to use this equation to numerically calculate the expansion in the Lamb-Dicke parameter $\eta$ to order $2m$. Our simulations are conducted with an expansion in the Lamb-Dicke parameter to at least order 12, with at least 17 number states for the motion included.

\subsection{Experimental parameters}
Simulations are conducted using measured rates and characteristic repump time for the scaling of Hamiltonian and dissipation terms. For $-$315 GHz Raman detuning these are:
\begin{align}
    \Omega_{bq} = 2\pi\times 6.56 \:\mathrm{kHz},\\
    \Omega_{ba} = 2\pi\times 10.03 \:\mathrm{kHz},\\
    \Omega_c = 2\pi\times 4.08 \:\mathrm{kHz},\\
    t_{rep} = 34 \:\mathrm{\mu s},
\end{align}
where $t_{rep}$ is the depletion time constant of the auxiliary state $\ket{\mathrm{aux}}$. For the $-$450 GHz detuning case, the measured Rabi frequencies and repump time are:
\begin{align}
    \Omega_{bq} = 2\pi\times 3.43 \:\mathrm{kHz},\\
    \Omega_{ba} = 2\pi\times 5.50 \:\mathrm{kHz},\\
    \Omega_c = 2\pi\times 1.77 \:\mathrm{kHz},\\
    t_{rep} = 69.5 \:\mathrm{\mu s}.
\end{align}
In preparation for taking the data presented in the main text, these parameters were initialized to values based on Ref. \cite{Horn2018} and then adjusted from there to increase the singlet fidelity. The values given above were measured after singlet generation.

\subsection{Optimal parameters in the large-detuning limit}

We have calculated optimal interaction strengths in the large-detuning limit for a Lamb-Dicke parameter $\eta=0.257$ and found a maximum fidelity of 0.989. In addition to optimizing the Raman-beam polarizations, we optimize the Raman-beam power distribution. For fixed total Raman beam power $P_{tot}$, we parameterize the power distribution with $P_b$, $P_r=P_{tot}-P_b$, and $r_q$, where the latter specifies the fraction of the red-beam power that drives the qubit transition. Then the powers at the frequencies $\omega_b$, $\omega_{rq}$, and $\omega_{ra}$ are $P_b$, $r_q P_r$, and $(1-r_q)P_r$. Independent of the other parameters, the optimal red/blue distribution is $P_b=P_r=P_{tot}/2$ because the Rabi frequency of each stimulated-Raman sideband transition is proportional to the common factor $\sqrt{P_b P_r}$, which should simply be maximized. This condition is not met in the experiment; instead, $P_r\approx0.4 P_b$ in both cases due to the technical details of the generation of the Raman beams.

We find optimal polarizations and red-beam power distribution of $b_\pi = 0.59$, $r_+=0.88$, and $r_q=35.7$ \%. Under these conditions, $\Omega_{bq}/\Omega_{ba}=0.58$. For comparison, the measurement of this ratio for the $-$315 GHz ($-$450 GHz) Raman-detuning experiment was 0.65 (0.62).

The optimal values for the other interaction strengths are $\Omega_c/\Omega_{ba}=0.27$ and $t_{rep}\Omega_{ba}/\pi=0.22$. The corresponding values for the $-$315 GHz ($-$450 GHz) detuning experiment were 0.41 and 0.68 (0.32 and 0.76), respectively. Investigation of the optimal repumping strength in the case of finite detuning would be useful in the future.

\subsection{Deviation from optimal parameters}
\begin{figure}[]
    \begin{centering}
	\includegraphics[]{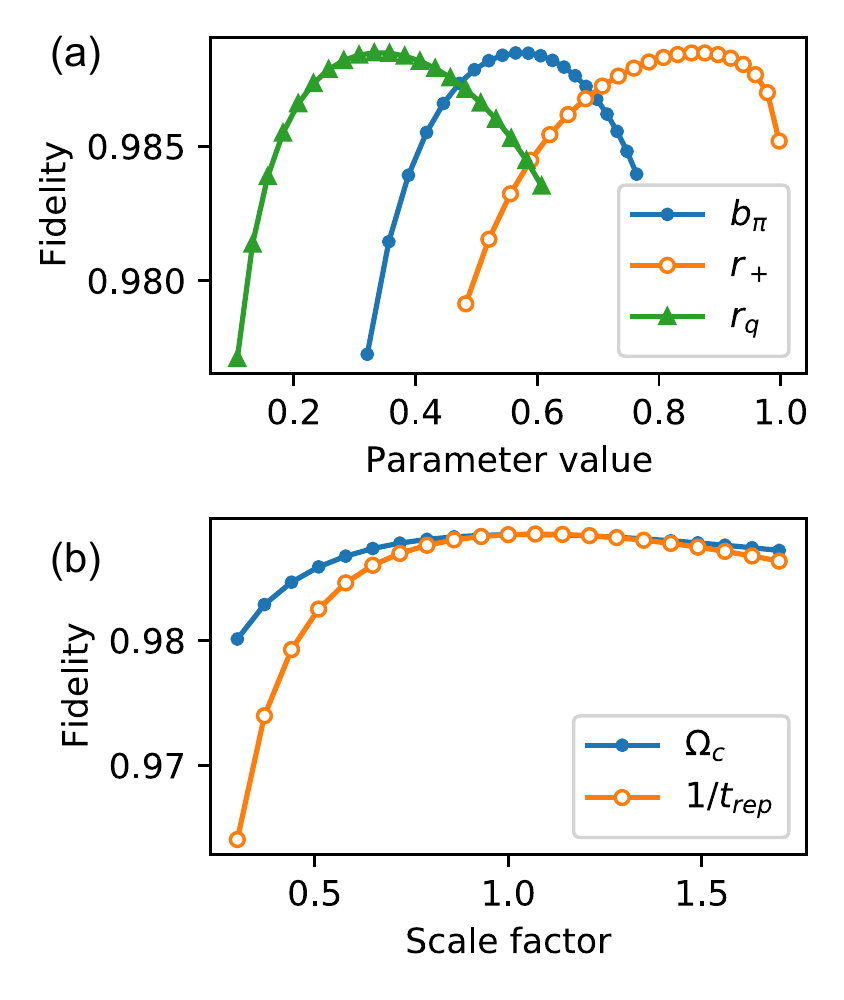}
	\caption{Investigation of sensitivity to deviation of parameters from their optimal values. (a)~Large-detuning-limit fidelity as a function of $b_\pi$, $r_+$, or $r_q$ with all other parameters held at their optimal values. (b)~Dependence of the fidelity on $\Omega_c$ or $t_{rep}$ when other parameters are optimal. Fidelity is plotted as a function of the factor by which the optimal value is scaled in the simulation. \label{fig:optdeviation}}
	\end{centering}
\end{figure}

To investigate the sensitivity of the protocol to deviations from the optimal parameters, we have calculated the fidelity in the large-detuning limit when one of $b_\pi$, $r_+$, $r_q$, $\Omega_c$, or $t_{rep}$ is away from its optimal value. We present the results in Fig. \ref{fig:optdeviation}.

\section{Calibration of the spacing between the ions}
It is important that the projection of the inter-ion separation vector ($\vec{r}_2 - \vec{r}_1$ as defined in the main text) onto the direction of the Raman beam $\Delta\vec{k}$ vector is $m+1/2$ (integer $m$) wavelengths of the $\Delta\vec{k}$ vector. This leads to $\phi=\Delta\vec{k}\cdot(\vec{r}_2-\vec{r_1})\mod 2\pi=\pi$ such that $H_{bq}^{(e)}$ coincides with $H_{bq}$, as described in the main text (see Eq. (4)).

We measure and calibrate $\phi$ using the following sequence \cite{Chou2011}: After we prepare the state $\ket{\downarrow\downarrow}$, a microwave $\pi/2$ pulse with randomized phase maps the qubits to a random unit vector on the equator of the Bloch sphere. We then apply a carrier $\pi/2$ pulse using the Raman beams, which evolves the qubits according to the Hamiltonian:
\begin{equation}
    H_{\mathrm{CR}}= e^{i\Phi}\frac{\hbar\Omega_{CR}}{2}\left(\ket{\uparrow}_1\bra{\downarrow}_1 + e^{i\phi}  \ket{\uparrow}_2\bra{\downarrow}_2\right) + H.c.
\end{equation}
The phase $\Phi = \Delta\vec{k}\cdot\vec{r}_1+\theta$ (as defined in the main text) fluctuates from shot to shot. The Rabi frequency of the interaction is $\Omega_{\mathrm{CR}}$, where $\mathrm{CR}$ denotes `Carrier Raman.' 

The carrier Raman pulse rotates the two qubits around two equatorial axes that have an angle $\phi$ between them. This angle is revealed by the final populations. Averaged over the randomized phase of the initial microwave $\pi/2$ pulse and the fluctuating phase $\Phi$, the populations are:
\begin{eqnarray}
P_{\downarrow\downarrow}+P_{\uparrow\uparrow} = \frac{1}{2}+\frac{1}{4}\cos\phi, \\
P_{\downarrow\uparrow} + P_{\uparrow\downarrow} = \frac{1}{2} - \frac{1}{4}\cos\phi. \label{eq:anticorr}
\end{eqnarray}
In practice, we fix $\phi=\pi$ by measuring the state parity as a function of a shim voltage applied to a pair of DC control electrodes that adjusts the curvature of the axial potential, and then applying the shim that maximizes the probability of an anti-correlated final state (Eq. \eqref{eq:anticorr}). In our case, this fixes the ion spacing to 16.5$\tilde{\lambda}$, where $\tilde{\lambda} = \lambda/\sqrt{2}$ is the wavelength of the interference pattern between the Raman laser beams with wavelength $\lambda\sim313$ nm.

\begin{figure}[]
    \begin{centering}
	\includegraphics[]{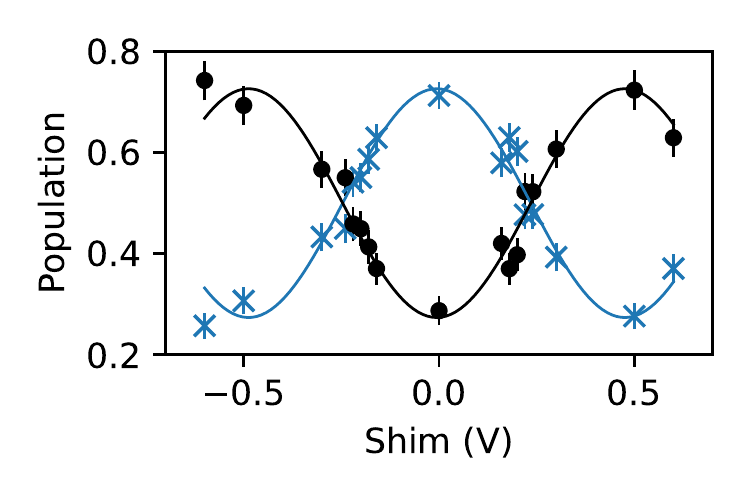}
	\caption{Typical results for the calibration of the phase $\phi$ discussed in the main text. Data and fits corresponding to the $+1$-parity population $P_{\downarrow\downarrow}+P_{\uparrow\uparrow}$ (black dots and line) and $-1$-parity population $P_{\downarrow\uparrow}+P_{\uparrow\downarrow}$ (blue crosses and line) are shown as a function of a static shim voltage applied to a pair of trap electrodes to adjust the curvature of the axial potential. Error bars indicate 68 \% confidence intervals. \label{fig:ion_spacing}}
	\end{centering}
\end{figure}

Fig. \ref{fig:ion_spacing} shows typical results for the measurement described above. The fits to the data shown share contrast and phase parameters. Typical uncertainty in the calibration of the phase $\phi$ is $\sim$0.05 rad. For a phase of $\phi=\pi\pm0.05$, and neglecting all other error sources, the simulated singlet fidelity is 0.993. A $\phi$ calibration error is included in the simulations for $-$315 GHz detuning as described in the main text, and $\phi=\pi$ is used in the simulations for $-$450 GHz detuning.

\section{Evaluation of other error sources}

\begin{figure*}[]
    \begin{centering}
	\includegraphics[]{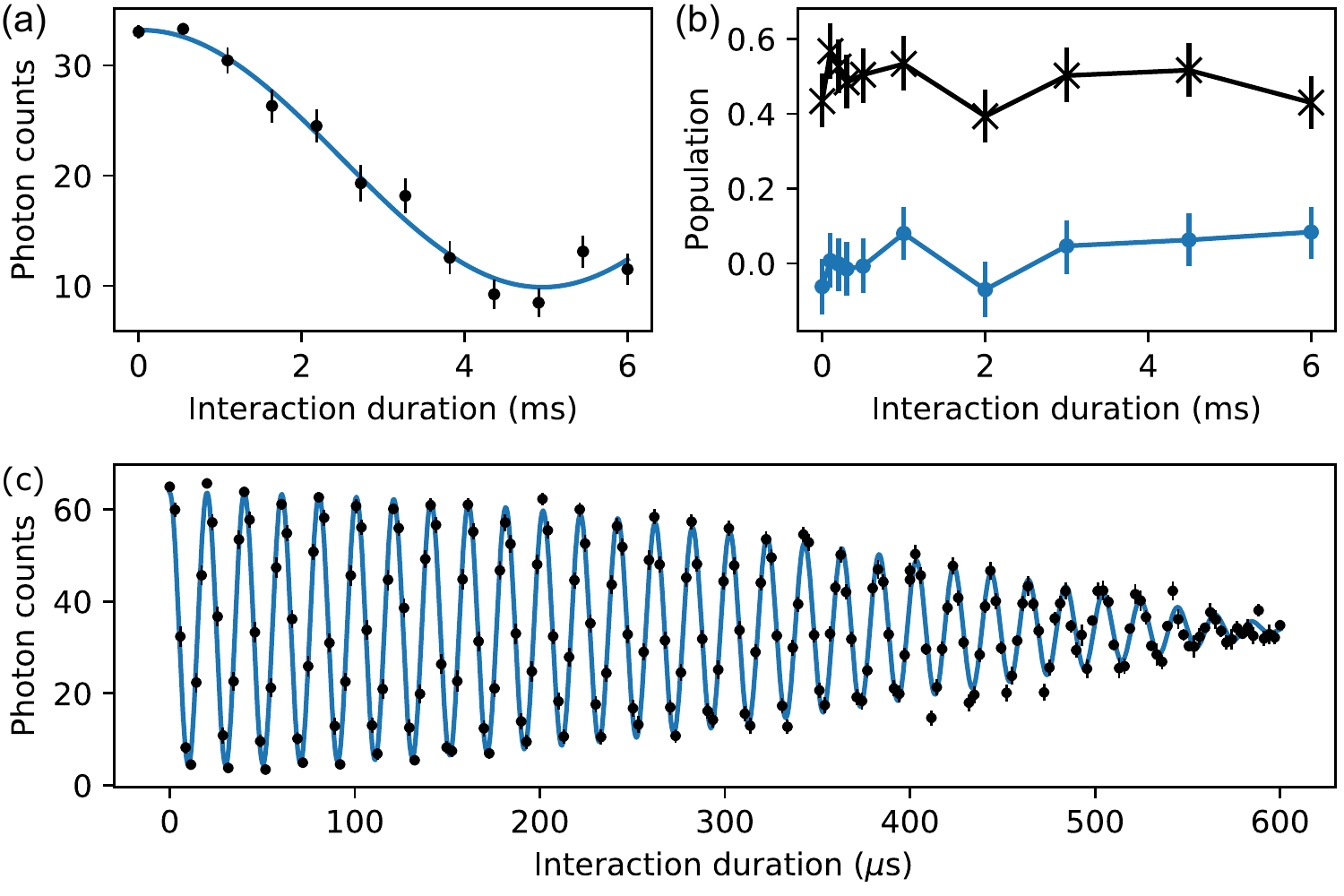}
	\caption{Measurements of error sources. (a) A measurement with a single ion of the residual unitary coupling between the $\ket{\downarrow}$ ($\sim$ 32 photon counts) and $\ket{\mathrm{aux}}$ ($\sim$ 3 photon counts) states. The reduced contrast of the Rabi oscillation is likely due to oscillations in the magnetic field direction at 60 Hz. (b) A measurement indicating negligible $\ket{T}\leftrightarrow\ket{S}$ coupling over 6 ms, with the Raman beams illuminating the ions. Shown are populations in the $\ket{T}$ state (black crosses) and the $\ket{S}$ state (blue dots) as a function of illumination duration after preparing $(\ket{\downarrow}_1+\ket{\uparrow}_1)\otimes(\ket{\downarrow}_2+\ket{\uparrow}_2)/2$ with 50~\% population in $\ket{T}$. The $\ket{T}\leftrightarrow\ket{S}$ coupling has a $\pi$ time of about 3 ms with the Raman beams off due to a residual magnetic field gradient, which the Raman beams are aligned to compensate. (c) Qubit-transition Rabi oscillations on two ions. From this measurement and the fit (blue) we estimate a Rabi-frequency difference of $\sim$1.7 \% between the two ions. The coherence time of this field-sensitive qubit is $\gtrsim 1$ ms. Error bars indicate 68 \% confidence intervals on all data points.\label{fig:other_errs}}
	\end{centering}
\end{figure*}

In our implementation we minimize the effects of error sources other than photon-scattering error and quantify their contributions. Besides a possible $\phi$ calibration error as discussed above, we have identified only one other important error source, which is the residual unitary $\ket{\downarrow}\leftrightarrow\ket{\mathrm{aux}}$ coupling described in the main text. Here we present our measurements of this error source and two others: qubit frequency differences and differences in the qubit-sideband Rabi frequencies $\Omega_{bq,1}$ and $\Omega_{bq,2}$ on the two ions. The results of measurements used to characterize these three error sources are shown in Fig.~\ref{fig:other_errs}.

Fig. \ref{fig:other_errs}a shows a measurement of the residual $\ket{\downarrow}\leftrightarrow\ket{\mathrm{aux}}$ coupling, which has a $\pi$ time of about 5 ms. The coupling is measured here with $-$315 GHz Raman detuning. The coupling is included in the simulations presented in the main text, where it decreases the singlet fidelity by $\sim$ 0.008. To include the coupling in the simulation for the case of $-$450 GHz Raman detuning, the strength of the coupling is scaled appropriately from the $-315$ GHz-detuning measurement.

Fig. \ref{fig:other_errs}b shows the results of a measurement used to quantify differences in the qubit frequencies of the two ions. In this measurement, the qubits are initialized in the state $\ket{\downarrow\downarrow}$, and then a $\pi/2$ pulse is applied to create the state $(\ket{\downarrow}_1+\ket{\uparrow}_1)\otimes(\ket{\downarrow}_2+\ket{\uparrow}_2)/2$ with 50~\% population in the $\ket{T}=\left(\ket{\uparrow\downarrow}+\ket{\downarrow\uparrow}\right)/\sqrt{2}$ state. Qubit frequency differences lead to a $\ket{T}\leftrightarrow\ket{S}$ coupling, which we detect by measuring the $\ket{S}$ and $\ket{T}$ populations as described in the main text after a delay of variable duration. We can make this measurement with the Raman beams on or off to quantify the contribution of differential ac Stark shifts from these beams to the $\ket{T}\leftrightarrow\ket{S}$ coupling. We first minimize magnetic field gradients by minimizing the $\ket{T}\leftrightarrow\ket{S}$ coupling with the Raman beams off by making adjustments to the current through several magnetic field coils near the ion trap. Then we minimize the $\ket{T}\leftrightarrow\ket{S}$ coupling with the Raman beams on through careful beam alignment. Other sources (e.g. the repumper laser and ac Zeeman shifts from the microwaves that implement the carrier drive $H_c$) do not contribute significantly to qubit frequency differences. We are able to reduce the coupling so that it has only a small effect over 6 ms, as shown in Fig. \ref{fig:other_errs}b.

Fig. \ref{fig:other_errs}c shows qubit Rabi oscillations on two ions, with loss of contrast coming primarily from differences in the Rabi frequencies on the two ions. The qubit coherence time of $\gtrsim$ 1 ms also contributes to loss of coherence, but a coherence revival beyond 600 $\mu$s allows estimation of $\sim$1.7~\% Rabi-frequency difference between the two ions. This difference arises because of a residual magnetic field gradient that is compensated for by adjusting the pointing of the Raman beams and thereby introducing small differential ac Stark shifts. As a result, the $\ket{T}\leftrightarrow\ket{S}$ coupling is minimized at the cost of introducing slight Rabi-frequency differences. The singlet fidelity calculated including only this error source is greater than 0.999.

\section{Two-qubit basis-state population measurements}
For each Raman detuning $\Delta$ we measure two-qubit basis-state populations as a function of dissipative interaction duration. In each case a duration $t_{max}$ is chosen. The ions are prepared approximately in $\ket{\downarrow\downarrow, n=0}$ using optical pumping, Doppler cooling, and sideband cooling, and then the light fields driving the dissipative dynamics are applied for duration $t$. The analysis pulses are applied as described in the main text. Then, after a delay of duration $t_{max}-t$, whose purpose is to hold constant the duty cycle for the detection laser (which is much higher power than the repumper that participates in the dissipative dynamics), fluorescence detection is performed.

The above procedure yields a set of observed photon counts $c_j(t,A)$, where $j$ indexes over repetitions of interaction duration $t$ and analysis condition $A$. For each data point shown in Fig. 2 in the main text, at least two hundred repetitions were performed for each analysis condition, and three times as many repetitions were performed for the $\pi/2$ pulse with randomized phase as for the other two conditions. 

This full set of counts is used to determine parameters for a detection model using maximum-likelihood estimation. The four model parameters are $C_{(0,1)}$ and $\mathrm{Pr}_{(0,1)}$. The former describe the background counts and the average number of counts detected from a bright ion, and the latter give the probabilities for observing zero or one ion bright. These two parameters also determine the expected number of counts $C_2$ for two bright ions and the corresponding observation probability $\mathrm{Pr}_2=1-\mathrm{Pr}_0-\mathrm{Pr}_1$. The observed counts are assumed to be Poisson-distributed about these calibrated parameters. Maximum-likelihood estimation then uses the parameters $C_n$ and the corresponding Poisson distributions to estimate the populations $P_{n,A}(t)$ as defined in the main text, from which the basis-state populations are determined according to Eqs. (5)-(8) in the main text.

Uncertainties in the reported populations are obtained using bootstrapping. For each analysis condition and interaction time, a data set of the same size as the experimental data is obtained by sampling from the data with replacement. The full analysis procedure is then run on this batch of synthetic data, yielding bootstrapped basis-state populations at each interaction time. After repeating this process $N=$10,000 times, we obtain bias-corrected (so-called `BC$_a$' with acceleration parameter $a=0$) confidence intervals on all reported quantities from the distribution of the corresponding bootstrapped parameters \cite{Efron1987}.

\bibliography{SupplementaryInformation}